\documentclass{nature}
\usepackage{graphicx}
\usepackage{amsmath}
\usepackage{epstopdf}
\usepackage{float}
\usepackage{xcolor}
\usepackage{ccaption}
\makeatletter
\let\saved@includegraphics\includegraphics
\AtBeginDocument{\let\includegraphics\saved@includegraphics}
\renewenvironment*{figure}{\@float{figure}}{\end@float}
\makeatother

\newcommand{\beqa}{\begin{eqnarray}}
\newcommand{\eeqa}{\end{eqnarray}}

\newcommand{\up}{{\left|\uparrow\right.}}
\newcommand{\down}{{\left|\downarrow\right.}}

\newcommand{\ldown}{{\left.\downarrow\right|}}

\title{Observation of parity-time symmetry breaking transitions in a dissipative Floquet system of ultracold atoms}
\author{Jiaming Li$^{1}$, Andrew K. Harter$^2$, Ji Liu$^2$, Leonardo de Melo$^{1,2}$, Yogesh N. Joglekar$^2$, \& Le Luo$^{1,2}$}

\begin{document}
\maketitle

\begin{affiliations}
\item School of Physics and Astronomy, Sun Yat-sen University, Zhuhai 519082, China
\item Department of Physics, Indiana University Purdue University Indianapolis (IUPUI), Indianapolis, Indiana 46202, USA
\end{affiliations}


\begin{abstract}
Open physical systems with balanced loss and gain, described by
non-Hermitian parity-time ($\mathcal{PT}$) reflection symmetric
Hamiltonians, exhibit a transition which could engenders modes that
exponentially decay or grow with time and thus spontaneously breaks
the $\mathcal{PT}$-symmetry. Such $\mathcal{PT}$-symmetry breaking
transitions have attracted many interests because of their
extraordinary behaviors and functionalities absent in closed
systems. Here we report on the observation of
$\mathcal{PT}$-symmetry breaking transitions by engineering
time-periodic dissipation and coupling, which are realized through
state-dependent atom loss in an optical dipole trap of ultracold
$^6$Li atoms. Comparing with a single transition appearing for
static dissipation, the time-periodic counterpart undergoes
$\mathcal{PT}$-symmetry breaking and restoring transitions at
vanishingly small dissipation strength in both single and
multiphoton transition domains, revealing rich phase structures
associated to a Floquet open system. The results enable ultracold
atoms to be a versatile tool for studying $\mathcal{PT}$-symmetric
quantum systems.
\end{abstract}
\maketitle


\section{Introduction}
A non-Hermitian parity-time reflection symmetric
($\mathcal{PT}$-symmetric) Hamiltonian, that is invariant under
combined parity ($\mathcal{P}$) and time-reversal ($\mathcal{T}$)
operations, has been considered as a natural extension of the
conventional Hermitian quantum theory to describe an open quantum
system with balanced loss and
gain~\cite{Bender1998,ChristOL2007,BenderEuroPhys2016}.
$\mathcal{PT}$-symmetric Hamiltonians exhibit many interesting
behaviors~\cite{Lin11PRL106-213901,Feng11S333-729,Feng13NM12-108,BPeng14S346-328,Hodaei14S346-975,Feng14S346-972},
in which a key property is $\mathcal{PT}$-symmetry breaking
transitions that occur at an exceptional
point~\cite{Kato1966,WDHeiss04JPhysA} -- a point in the parameter
space where two resonant modes of the Hamiltonian become degenerate.
A number of seminal studies~\cite{Bender1998,Bender2007} have shown
that the eigenvalues of the Hamiltonian are real in one side of the
transition, allowing the $\mathcal{PT}$-symmetric (PTS) phase, while
complex eigenvalues appear in the other side with the
$\mathcal{PT}$-symmetric broken (PTSB) phase. In recent years,
$\mathcal{PT}$-symmetric Hamiltonians have been realized in balanced
gain and loss systems with various setups, such as mechanical
oscillators~\cite{Bender13AJP81-173}, optical
waveguides~\cite{Ruter10NP6-192,Regensburger2012}, optical
resonators~\cite{Lee2009}, microwave cavities~\cite{Dembowski2001},
lasers~\cite{Brandstetter2014}, and optomechanical
systems~\cite{Xu2016}, or in a state dependent pure lossy system in
which the lossy Hamiltonian $H'$ could be mapped to a
$\mathcal{PT}$-symmetric Hamiltonian $H_{\text{PT}}$ for passive
$\mathcal{PT}$-symmetry
breaking~\cite{Lee2015,Christ09PRL103-093902,Xiao2017}.

$\mathcal{PT}$-transitions can be induced either by increasing the
strength of dissipation or by tuning the periodicity of the
dissipation, known as static or Floquet method respectively. By
driving the system passing the exceptional point, it is predicted
that $\mathcal{PT}$-transitions can reduce the overall dissipation
of the system~\cite{Joglekar2014, Lee2015, Rabi2004} and allow
topological structures around the exceptional
point~\cite{Xu2016,Doppler2016}. Floquet method is particular
interesting because time-periodic modulation can break the
continuous time translation symmetry, providing an enriched phase
diagram with many fascinating features~\cite{Lee2015}.

Here we present an experimental study of $\mathcal{PT}$-symmetry
breaking transitions induced by time-periodic dissipation or
coupling in a two-spin system of ultracold atoms. Our experimental
results verify that $\mathcal{PT}$-symmetry breaking and restoring
transitions can occur by tuning either dissipative or coupling
frequency even at vanishingly small dissipation strength. We further
map the Floquet $\mathcal{PT}$-phase diagrams by tracing the atom
loss of each spin state, and observe the multiphoton resonances and
the power broadening associated to the PTSB phase.

\section{Results}
\subsection{$\mathcal{PT}$ transition with static dissipation.}
We prepare a noninteracting Fermi gas of $^6$Li atoms at the two
lowest $^2\text{S}_{1/2}$ hyperfine levels~\cite{JL16,ohara02a}, labeled as
$\up\rangle$ and $\down\rangle$. These two-spin states are coupled
by a radio-frequency (RF) field with a coupling strength of $J$. A
resonant optical beam is used to excite the atoms from
$\down\rangle$ to the 2P excited state $\left|e\right\rangle$ and
generates the atom loss in $\down\rangle$ with a rate of $\Gamma$
(Fig.~\ref{fig1}a). The Hamiltonian for this dissipative two-spin
system is given by
\begin{equation}
\label{eq:hmain}
H(t)=J\sigma_{\text{x}}-i\Gamma(t)\down\rangle\langle\ldown=-i\Gamma(t)/2\,\mathbf{I}
+ H_{\text{PT}}(t)
\end{equation}
where
$H_{\text{PT}}(t)=J\sigma_{\text{x}}+i\Gamma(t)\sigma_{\text{z}}/2$
is a $\mathcal{PT}$-symmetric Hamiltonian, and $\mathbf{I}$ is the
unit matrix. The system is prepared with all atoms in $\up\rangle$,
and evolves for a time of $t$. Then the in-trap atom numbers,
$n'_\uparrow(t)$ and $n'_\downarrow(t)$, are measured by the
double-shot absorption imaging of the two spin states, giving the
total atom number $n'(t)=n'_\uparrow+n'_\downarrow$. We map $n'(t)$
to a scaled, normalized atom number $n(t)$ associated to
$H_{\text{PT}}$, and then use $n(t)$ to characterize the
$\mathcal{PT}$-transitions [See Methods].

For static dissipation, $\Gamma (t)$ is a constant value of
$\Gamma_0$. When $\Gamma_0/J<2$, the eigenvalues of $H_{\text{PT}}$
are real values of $\pm\sqrt{J^2-\Gamma_0^2/4}$, and $n(t)$
oscillates at frequency $\pi/\sqrt{J^2-\Gamma_0^2/4}$. The
$\mathcal{PT}$ transition occurs at $\Gamma_0/J=2$ where the
oscillation period diverges. When $\Gamma_0/J>2$, the eigenvalues of
$H_{\text{PT}}$ become complex numbers, and one of the eigenmode
exponentially grows [See Supplementary Note 1]. These predictions
are verified in our experiments [See Supplementary Figure 1], and
the measured exceptional point agrees with the theoretical model
very well [See Supplementary Figure 2].

The static dissipation experiment is related to the previous quantum
zeno effect (QZE) experiments of ultracold atoms with strong-loss
induced
measurement~\cite{KetterleZeno,RaizenZeno,Zhu2014,PatilZeno}. In
those experiments, QZE refers to the reduction of the rate of
transferring from one state to a second state by the projection
measurement of the second state. Due to a strong-loss induced
irreversible measurement, the reverse-transfer probability from the
second state to the first one is treated as zero as well as the
occupation of the second state. However, in our dissipation
experiment, the transfer probability from the second to the first
level is nonzero, and the PTSB phase refers to the slow-down of the
decay of the total atom number. Thus, the results cannot be
explained purely in terms of QZE, except for the limit of an
extremely strong dissipation case, in which the strong atom loss can
be treated as an irreversible projection measurement of the second
level.

\subsection{Observation of the $\mathcal{PT}$-transitions with
time-periodic driving.} Floquet method enriches the phase diagram of
a $\mathcal{PT}$-symmetric system by periodically modulating
Hamiltonian $H(t)=H(t+T)$. Previously, the extraordinary structure
of the phase diagram has been theoretically
predicted~\cite{Joglekar2014, Lee2015}, but has never been verified
experimentally due to the difficulty of precisely controlling the
time-dependent dissipation. In our experiment, the optical and RF
field provide versatile tools to manipulate the atom loss and
coupling of spin levels, so that two types of Floquet Hamiltonians
could be implemented: spin-dependent time-periodic dissipation and
time-periodic coupling between two spins.

\begin{figure}[htbp]
\includegraphics[width=0.6\columnwidth]{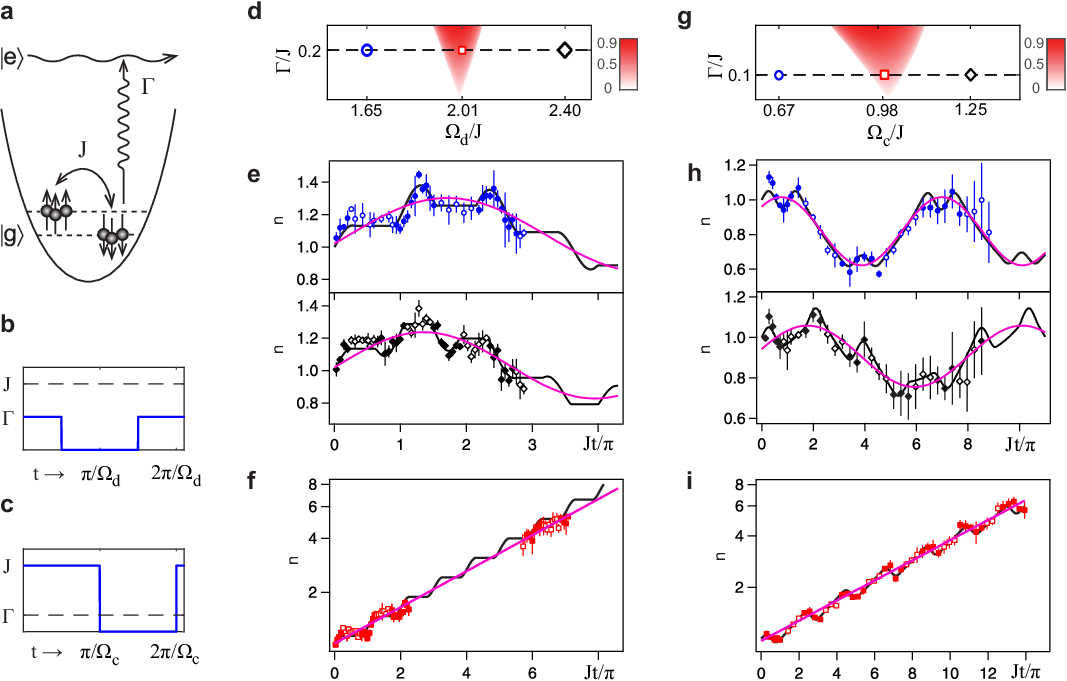}
\caption{The parity-time transitions induced by time-periodic
modulations. \textbf{a} Experimental setup. An RF field is used to
couple the two-spin states. A resonant optical beam is used to
generate spin-dependent dissipation (atom loss) in the
$\down\rangle$ state. \textbf{b} $\Gamma(t)$ for time-periodic
dissipation. \textbf{c} $J(t)$ for time-periodic coupling.
\textbf{d-f} Time-periodic dissipation: \textbf{d} The phase diagram
near the primary resonance. The red color region represents the
$\mathcal{PT}$-symmetric broken (PTSB) phase with
$(|\mu_+|-|\mu_-|)/(|\mu_+|+|\mu_-|)$ as the value of the color
density, as the following figures. \textbf{e} $n(t)$ of the
$\mathcal{PT}$-symmetric (PTS) phase at $\Omega_{\text{d}}/J=1.65$
(blue circles) and $\Omega_{\text{d}}/J=2.40$ (black diamonds).
\textbf{f} $n(t)$ of the PTSB phase at $\Omega_{\text{d}}/J=2.01$
(red boxes). \textbf{g-i} Time-periodic coupling: \textbf{g} The
phase diagram near the primary resonance. \textbf{h} $n(t)$ of the
PTS phase at $\Omega_{\text{d}}/J=0.67$ (blue circles) and
$\Omega_{\text{d}}/J=1.25$ (black diamonds). \textbf{i} $n(t)$ of
the PTSB phase at $\Omega_{\text{d}}/J=0.99$ (red boxes). In
\textbf{e} and \textbf{f} (\textbf{h} and \textbf{i}), the data with
solid shapes corresponds with the dissipation (coupling) on, and the
data with empty shapes are with the dissipation (coupling) off. In
all figures, the black curves are the numerical simulation without
free parameter, and the pink curves are the sinusoidal (for PTS) or
exponential (for PTSB) fitting. $J=\pi\times2.15$ kHz for all data
presented in this paper if not mentioned. The error bars are the
standard deviation of the measurements.\label{fig1}}
\end{figure}

We first study time-periodic dissipation, in which a square-wave
resonant beam is applied to generate time-dependent dissipation of
the atoms in an optical trap. The coupling strength $J$ is fixed and
the dissipation strength is modulated between $\Gamma$ and 0 with a
frequency of $\Omega_{\text{d}}$ (Fig~\ref{fig1}.b). In contrast with the
static dissipation, $\mathcal{PT}$-transitions under time-periodic
dissipation depends on the modulation frequency and can occur at
vanishingly small dissipation strength with infinite numbers of the
resonance peaks [See the Supplementary Note 2]. The primary
resonance peak of the $\mathcal{PT}$-transition appears
$\Omega_{\text{d}}/J=2$, where the transition behavior of $n(t)$ in the weak
dissipation limit $\Gamma/J=0.2\ll2$ is shown in Fig~\ref{fig1}.d-f.
When $\Omega_{\text{d}}/J$ is tuned to the PTSB phase, $n(t)$ increases
exponentially, in contrast with the PTS phase where $n(t)$ exhibits
bounded oscillation $n(t)\propto \sin[(\Omega_{\text{d}}-2J)t]$.

In the above cases, the PTSB phases have been observed even when the
eigenvalues of $H(t)$ are real all the time. Such
$\mathcal{PT}$-symmetry breaking can be determined by the
non-unitary time evolution operator $G_{\text{PT}}$ [See Methods], which
has two eigenvalues $\mu_\pm\propto e^{-i\epsilon_\pm t}$.
$\epsilon_\pm$ is the quasienergies of the effective Floquet
Hamiltonian [See Supplementary Note 3] . If the magnitude of
$\mu_\pm$ are equal, $e^{i\epsilon_\pm T}$ is a pure phase factor
and $\epsilon_\pm$ must be the real numbers, which indicate a PTS
phase. On the contrary, the unequal magnitude of $\mu_\pm$ denote
the complex values of $\epsilon_\pm$ representing a PTSB phase.

For time-periodic coupling in the weak dissipation limit, $\Gamma/J$
is constant and $J(t)$ is modulated at the frequency
$\Omega_{\text{c}}$ (Fig.~\ref{fig1}c) and the the primary resonance
peak of the PTSB phase is at $\Omega_{\text{c}}/J=1$
(Fig.~\ref{fig1}g). When $\Omega_{\text{c}}/J$ is tuned to the
primary resonance region, $n(t)$ shows the similar behavior as
time-periodic dissipation, where the exponential increase of $n(t)$
appears (Fig.~\ref{fig1}i), while the PTS phase exhibits the bound
oscillation which could be parameterized by $n(t)\propto
\sin[(\Omega_{\text{c}}-J)t]$ in the weak dissipation limit
(Fig.~\ref{fig1}h). The measurements of that primary resonance of
the PTSB phase verify that $\mathcal{PT}$-symmetry transitions can
happen with an arbitrary small dissipation under time-periodic
driving. Furthermore, there exist infinite numbers of transitions
induced by multiphoton resonances which are investigated as follows.

\subsection{Multiphoton resonances with time-periodic dissipation.}
For $\mathcal{PT}$-transitions with time-periodic dissipation, there
exist infinite numbers of the PTSB phases induced by multiphoton
process in a non-Hermitian Rabi model~\cite{Lee2015}. Their widths
have been predicted to decrease with the index number of the the
resonances. For a square-wave modulation, the widths of the PTSB
phases in the weak dissipation limit are
\begin{equation}
\label{eq:Gammawidth}
\delta\Omega_{\text{d}}(\Gamma,\Omega_{\text{n}})=\frac{\Gamma}{\pi}\left(\frac{\Omega_{\text{n}}}{2J}\right)^2,
\end{equation}
where $\Omega_{\text{n}}=2J/n$ is the resonance peak under zero
dissipation with $n$ as the odd number 1,3,5.... $\Gamma$ is the
magnitude of the square-wave dissipation [See Supplementary Note 2].
Fig.~\ref{fig2} show the broadening of the PTSB phases for one-
(primary), three-, and five-photon resonances. To measure the width
of the resonances, the residual atom number $n(t_{\text{f}},\Omega)$
is probed at a fixed time point $t_{\text{f}}$ for various
modulation frequencies $\Omega$ [See Supplementary Note 4]. It is
noted that, for the purpose of mapping the phase diagram, it is
ideal to choose $t_{\text{f}}$ as large as possible so that
$n(t_{\text{f}})$ can reflect the long-term dynamics. However,
because we map a pure lossy system to a $\mathcal{PT}$-symmetric
Hamiltonian, $t_{\text{f}}$ must be remained in a finite range for
the reasonable signal-to-noise ratio of the unscaled atom number
$n'(t)$ [See Supplementary Note 5]. In our experiment, we choose
$t_{\text{f}}$ to be larger than several oscillation periods so that
$n(t_{\text{f}},\Omega)$ can present the trend of increasing in the
PTSB phase. As shown in Fig.~\ref{fig2}a, the half width at half
maximum (HWHM) of the primary resonance is proportional to the
strength of time-periodic dissipation. Such behavior is the
non-Hermitian analog of the resonance broadening induced by the
Bloch-Siegert shifts of a strong driving Hermitian system. The width
of the residual atom number also depends on the probe time and gets
narrower for the longer probing times, which approaches the width of
the PTSB phase predicted by theoretical calculations. For the finite
probe time, the width is a qualitatively measure of the dependence
of the PTSB regime on the dissipation strength [See Supplementary
Note 4].

Comparing with the Hermitian system where the multiphoton resonance
is difficult to be observed at the weak driving limit, in a
non-Hermitian system, the time-periodic dissipation significantly
broadens the width of the PTSB phase so that the multiphoton
resonance could be observed clearly with the weak dissipation. The
widths of the three-(Fig.~\ref{fig2}b), five- (Fig.~\ref{fig2}d)
photon resonances, agree with the theoretical phase diagram very
well. At the exact resonant frequencies, the exponentially increase
of $n(t)$ with a very small dissipation strength are recorded in
Fig.~\ref{fig2}c (three-photon) and Fig.~\ref{fig2}e (five-photon)
manifesting the PTSB phase.

\subsection{Multiphoton resonances with time-periodic coupling.}
The phase diagram of time-periodic coupling is studied by modulating
the coupling between the two-spin states. The resonance widths of
the PTSB phases are given by
\begin{equation}
\label{eq:FCwidth} \delta\Omega_{\text{c}}(\Gamma,\Omega_{\text{n}})=\Gamma
\frac{\Omega_{\text{n}}}{2J}\,\,\,,
\end{equation}
where $\Omega_{\text{n}}=2J/n$ is the resonant peak with the even
integer $n=2,4,6...$. Eq.~\ref{eq:FCwidth} indicates that the PTSB
phases of the time-periodic coupling have the wider width than that
of time-periodic dissipation, which scales with the multiphoton
index number $n$ as $1/n$ instead of $1/n^2$ for time-periodic
dissipation [See Supplementary Note 2]. The first four multiphoton
resonances are shown in Fig.~\ref{fig3}a, where the widths of the
PTSB phases increase with dissipation. The increasing of $n(t)$ at
the resonance frequencies are fitted by the exponential curves in
Fig.\ref{fig3}b to verify the PTSB phase.

It is interesting to find that, the resonant peaks $\Omega_{\text{n}}=2J/n$
of the time-periodic dissipation have odd integers $n=1,3,5...$, but
the time-periodic coupling ones have even integers $n=2,4,6...$. The
odd or even rule can be explained by a simple picture. For example,
in the time-periodic coupling case of the weak loss limit, all atoms
are initially in the lossless (up) state and the coupling is turned
on for $n\pi/(2J)$. If n is even, then the coupling is on exactly
for the time of multiple $2\pi$ Rabi pulses such that all atoms are
back to the up state, and will remain in this state for the next
half cycle of coupling-off. During this half-cycle, no atoms are
lost so that the scaled total atom number increase because the
scaling assumes equal loss in both spin states. Overall, the system
spends more time in the lossless (up) state when $n$ is even
numbers. This is not the case for n to be odd, where on average the
system spends the same amount of time in both up and down states
because the coupling is turned on for the odd numbers of $\pi$
pulse.

The similar picture is also applied to the time-periodic dissipation
case, where the atom loss is turned on for a time duration of
$n\pi/(2J)$ with $n$ being odd numbers. This amount of time is odd
numbers of $\pi$ Rabi pulse. With the proper choice of phase, the
atom loss can only present when the majority of the atoms are in the
lossless (loss) state so that the scaled total atom number increase
(decrease) exponentially. Either increasing or decreasing depends on
the phase of the dissipation, which corresponds the two eigenstates
in the PTSB phase. In the experiments, we usually optimize the phase
of the time-periodic modulation to obtain the strongest signal,
resulting from ensuring the largest overlap between the initial
state and the slowly-decaying eigenmode of the corresponding Floquet
Hamiltonian. For more general initial states, such as a balanced
mixture of up and down state, as long as there is a nonzero overlap
between the initial state and the slow mode, the
$\mathcal{PT}$-symmetry breaking signatures of the slow-decaying
will be visible in the long-time limit.

\section{Discussion}
The laser-cooled ultracold atoms provide a clean and well
controllable platform for studying $\mathcal{PT}$-symmetric
Hamiltonians. Previously, phase transitions observed in cold atom
systems are usually driven by tunable interparticle interactions
and, in principle, occur only in the thermodynamic limit. However,
$\mathcal{PT}$-symmetric breaking transitions, in contrast, can
occur in a single two-level system with localized loss. The fate of
the former transitions in the presence of such a loss has not been
fully understood, as is the fate of latter transition in the
presence of interparticle interactions. Investigating the interplay
between these two classes of transitions will require quantum
simulators with tunable interparticle interactions and engineered
state-dependent dissipation, both of which can be realized with
certain species of ultracold atoms, such as fermionic $^6$Li atoms
used in this experiment. As a starting point of this route, we apply
state-dependent dissipation to ultracold $^6$Li atoms to study
$\mathcal{PT}$-transitions in a two-level system. With the
advantages of modulating the resonant optical and RF field as the
versatile tools for time-periodic driving, we could manipulate the
atom loss as well as the coupling of spin levels, and experimentally
map both time-periodic dissipation and coupling. The phase diagram
of the static dissipation and time-periodic dissipation (coupling)
are explored by tracing the time-evolution of the atoms. While the
single exceptional point under static dissipation is determined as
usual, our results verify remarkably rich phase diagrams with
multiple Floquet $\mathcal{PT}$-transitions associated to
time-periodic driving. It is shown that the PTSB phases can be
induced by judiciously selected temporal profiles of state-dependent
dissipation or coupling with vanishingly small strength of the
dissipation. The multiphoton resonant structure of
$\mathcal{PT}$-transitions are demonstrated. Such Floquet method
thus provide an experimental platform to study time-dependent
$\mathcal{PT}$-symmetric Hamiltonians.

Our system has potential to be extended to more complex situations:
one is to study the topological phenomena associated to
non-Hermitian Hamiltonian and the other is to explore an interacting
system with a vanishingly small, time-modulated dissipation. For the
formal one, if we use the unresonant RF pulses to couple the spin
levels, a detuning term will appear in the diagonal part of the
Hamiltonian, and we can adiabatically encircle the exceptional
points by changing the detuning and dissipation simultaneously to
observe the topological phenomena associated to the non-Hermitian
systems~\cite{Mailybaev2005,Uzdin2011,Wu2014,Gao2015,Milburn2015,Xu2016,Doppler2016,Hassan2017,Yoon2018}.
For the latter one, the interplay between the
$\mathcal{PT}$-transition and the BEC-BCS (Bose-Einstein condensate
to Bardeen-Cooper-Schrieffer pairing) crossover can be investigated
by sweeping the ultracold Fermi gas from the noninteracting limit
(presented here) to the unitary, strongly-interacting
limit~\cite{Luo07Entropy}. This approach, where a single-particle,
state-dependent loss is used in conjunction with strong
interparticle interactions, provides exciting opportunities to
explore physical phenomena in open many-body quantum systems.


\newpage
\section{Methods}
\subsection{Experimental System.} We prepare a dissipative two-level
system with a noninteracting Fermi gas. $^6$Li atoms are prepared in
the two lowest hyperfine states,
$\up\rangle\equiv|\text{F}=1/2,m_{\text{F}}=1/2\rangle$ and $\down\rangle\equiv
|\text{F}=1/2,m_{\text{F}}=-1/2\rangle$, in a magneto-optical trap. The precooled
atoms are then transferred into a crossed-beam optical dipole trap
made by a 100 Watt fiber laser. The bias magnetic field is swept to
330 G to implement an evaporative cooling~\cite{JL16}. The trap
potential is lowered to generate a final trap depth of 2.2 $\mu$K.
In order to null the interaction between the two hyperfine states,
the magnetic field is fast swept to 527.3 G, where the s-wave
scattering length of the $\up\rangle$ and $\down\rangle$ states is
zero~\cite{ohara02a}. The lifetime of the noninteracting Fermi gas
is about 20 seconds, which is three orders of magnitude longer than
our typical experimental time. So when the dissipative optical field
is absent, this noninteracting Fermi gas can be treated as a closed,
two-level quantum system. To prepare a single component Fermi gas in
the $\up\rangle$ state as the initial state, we apply a 5 ms optical
pulse with $-2\pi\times30$ MHz detuning from the
$\down\rangle\rightarrow\,^2\text{P}_{3/2}$ transition to blow away atoms
in the $\down\rangle$ state. We typically have about
$N=2.0\times10^5$ atoms in a pure $\up\rangle$ state at temperature
$T\approx$ 0.8 $\mu$K and $T/T_{\text{F}}\approx 0.5$ with $T_{\text{F}}$ is the Fermi
temperature.

To generate Rabi oscillation between the two spin states, we couple
them via an RF field with frequency $\omega$ and coupling strength
$J$. An optical beam resonant with the
$\down\rangle\rightarrow\,^2\text{P}_{3/2}$ transition is used to create
the number dissipation (atom loss) in the $\down\rangle$ state. The
resonant-photon recoil energy of 3.5 $\mu$K is approximately 50\%
larger than the trap depth, so the atom that absorb a photon will
leave the trap quickly, resulting a state-dependent atom loss. The
RF coupling strength $J$ is measured in the absence of the
dissipative optical field, while the atom-number loss rate $2\Gamma$
is measured in the absence of the RF coupling. Fig.~\ref{fig:sm12}a
shows the Rabi oscillation with Rabi frequency $2J$.
Fig.~\ref{fig:sm12}b shows the atom numbers
$n'_\downarrow(t)=n'_\downarrow(0)\exp(-2\Gamma t)$ with a constant
dissipative optical field that only couples the $\down\rangle$ state
to the continuum. These measurements are used to calibrate the
values of $J$ and $\Gamma$ for the dissipative two-state Rabi
system.

\subsection{Theoretical Model.} The dissipative two-state system is
described by a non-Hermitian Hamiltonian ($\hbar=1$)
\begin{equation}
\label{eq:TimeDept} H=+\frac{\omega_0}{2}\sigma_{\text{z}} - i\frac{\Gamma(t)}{2}(1-\sigma_{\text{z}})+ 2J\cos(\omega t) \sigma_{\text{x}},
\end{equation}
where $\omega_0=2\pi\times 75.6$ MHz is the hyperfine splitting at
527.3 G. When the RF driving is close to the resonance, that is
$\omega\approx\omega_0$, with the rotating wave approximation in the
interacting picture, $H(t)=- i\Gamma(t)/2+ H_{\text{PT}}(t)$, where the
non-Hermitian, $\mathcal{PT}$-symmetric Hamiltonian is given by
($\hbar=1$)
$H_{\text{PT}}=J\sigma_{\text{x}}+i\Gamma(t)\sigma_{\text{z}}/2=\mathcal{PT}H_{\text{PT}}\mathcal{PT}$,
where $\mathcal{P}=\sigma_{\text{x}}$ and $\mathcal{T}=*$ denotes complex
conjugation operation. Starting with an initial state
$|\psi(0)\rangle$, the decaying atom numbers for the two states are
given by $n'_{\sigma}(t)\equiv
|\langle\sigma|G'(t)|\psi(0)\rangle|^2$ where
\begin{equation}
\label{eq:g}
G'(t)=T\exp\left(-i\int_{0}^t H'(t')dt'\right),
\end{equation}
is the non-unitary time evolution operator obtained via the
time-ordered product. It is also useful to define scaled atom number
$n_\sigma(t)=|\langle\sigma|G_{\text{PT}}(t)|\psi(0)\rangle|^2$ where
$G_{\text{PT}}(t)$ is the corresponding time-evolution operator for
$H_{\text{PT}}(t)$. It follows that
$n_\sigma(t)=n'_\sigma(t)\times\exp(\int_0^t \Gamma(t')dt'/2)$. In a
$\mathcal{PT}$-symmetric system, the $\mathcal{PT}$-symmetric phase
is signaled by non-decaying, oscillatory $n_\sigma(t)$ and the
$\mathcal{PT}$-broken phase is signaled by an exponentially
increasing $n_\sigma(t)$.

\section{Data availability}
The data that support the findings of this study are available from
the corresponding author upon reasonable request.

\newpage

\begin{thebibliography}{40}
\bibitem{Bender1998}
{Bender, C.~M.\&Boettcher, S., Real spectra in non-{H}ermitian {H}amiltonians having
    $\mathcal{P}\mathcal{T}$ symmetry. \emph{Phys. Rev. Lett.}
    \textbf{80}, 5243--5246,  (1998).}

\bibitem{ChristOL2007}
{El-Ganainy, R., Makris, K.~G., Christodoulides, D.~N. \& Musslimani,Z.~H.
    Theory of coupled optical$\mathcal{P}\mathcal{T}$-symmetric structures. \emph{Opt. Lett.} \textbf{32},
    2632--2634 (2007).}

\bibitem{BenderEuroPhys2016}
{Bender, C.~M.$\mathcal{P}\mathcal{T}$ symmetry in quantum physics:
    From a mathematical curiosity to optical experiments. \emph{Europhys. News} \textbf{47}, 17--20 (2016).}

\bibitem{Lin11PRL106-213901}
{Lin, Z. et~al.Unidirectional invisibility induced by
    $\mathcal{P}\mathcal{T}$-symmetric periodic structures. \emph{Phys. Rev. Lett.}
    \textbf{106}, 213901 (2011).}

\bibitem{Feng11S333-729}
{Feng, L. et~al.Nonreciprocal light propagation in a silicon photonic
    circuit. \emph{Science} \textbf{333},
    729--733 (2011).}

\bibitem{Feng13NM12-108}
{Feng, L. et~al. Experimental demonstration of a unidirectional
    reflectionless parity-time metamaterial at optical frequencies. \emph{Nat. Materials}
    \textbf{12},108--113 ({2013).}

    \bibitem{BPeng14S346-328}
    {Peng, B. et~al. Loss-induced suppression and revival of lasing. \emph{Science} \textbf{346},
        328--332 (2014).}

    \bibitem{Hodaei14S346-975}
    {Hodaei, H., Miri, M.-A., Heinrich, M., Christodoulides, D.~N. \&
        Khajavikhan, M. Parity-time{\textendash}symmetric microring lasers. \emph{Science}  \textbf{346},
        975--978 (2014).}

    \bibitem{Feng14S346-972}
    {Feng, L., Wong, Z.~J., Ma,
        R.-M., Wang, Y.\& Zhang, X. Single-mode laser by parity-time symmetry breaking. \emph{Science} \textbf{346},
        972--975  (2014).

        \bibitem{Kato1966}
        {Kato, T. Perturbation theory for linear operators (Springer-Verlag, Berlin, New York,1966).}

        \bibitem{WDHeiss04JPhysA}
        {Heiss, W.~D. Exceptional points of non-{H}ermitian operators. \emph{J. of Phys. A: Math. and
            Gen.} \textbf{37} 2455 (2004).}

        \bibitem{Bender2007}
        {Bender, C.~M. Making sense of non-hermitian hamiltonians. \emph{Rept.Prog.Phys.}
            \textbf{70}, 947 (2007).}

\bibitem{Bender13AJP81-173}
        {Bender, C.~M., Berntson, B.~K., Parker, D. \&  Samuel, E.
            Observation of $\mathcal{P}\mathcal{T}$ phase
            transition in a simple mechanical system. \emph{Am. J. Phys.}}}
    \textbf{81}, 173--179
    (2013).}

\bibitem{Ruter10NP6-192}
{R\"{u}ter, C.~E. et~al. Observation of parity-time symmetry in
optics. \emph{Nat. Phys.}  \textbf{6},
    192--195 (2010).}

\bibitem{Regensburger2012}
{Regensburger, A. et~al. Parity-time synthetic photonic lattices.
\emph{Nature} \textbf{488},
    167--171 (2012).}

\bibitem{Lee2009}
{Lee, S.-B. et~al. Observation of an exceptional point in a chaotic
    optical microcavity. \emph{Phys. Rev. Lett.}
    \textbf{ 103}, 134101
    (2009).}

\bibitem{Dembowski2001}
{Dembowski, C. et~al. Experimental observation of the topological structure
    of exceptional points. \emph{Phys. Rev. Lett.}
    \textbf{86}, 787 (2001).}

\bibitem{Brandstetter2014}
{Brandstetter, M. et~al. Reversing the pump dependence of a laser at an
    exceptional point. \emph{Nat. Commun.}
    \textbf{ 5}, 192--195 (2014).}

\bibitem{Xu2016}
{Xu, H. Mason, D. Jiang,
    L. \& Harris, J. G.~E. Topological energy transfer in an optomechanical
    system with exceptional points. \emph{Nature}  \textbf{537},
    80 (2016).}

\bibitem{Lee2015}
{Lee, T.~E. \&  Joglekar, Y.~N. $\mathcal{P}\mathcal{T}$-symmetric {R}abi model:
    Perturbation theory. \emph{Phys. Rev. A} \textbf{92},
    042103 (2015).}

\bibitem{Christ09PRL103-093902}
{Guo, A. et~al. Observation of $\mathcal{P}\mathcal{T}$-symmetry
    breaking in complex optical potentials. \emph{Phys. Rev. Lett.}
    \textbf{ 103}, 093902 (2009).}

\bibitem{Xiao2017}
{Xiao, L. et~al. Observation of topological edge states in
    parity-time-symmetric quantum walks. \emph{Nat. Phys.} \textbf{13}, 1117 (2017).}

\bibitem{Joglekar2014}
{Joglekar, Y.~N., Marathe, R., Durganandini, P.  \&  Pathak, R.~K.
    $\mathcal{P}\mathcal{T}$ spectroscopy of the {R}abi
    problem. \emph{Phys. Rev. A } \textbf{ 90},
    040101 (2014).}

\bibitem{Rabi2004}
{Ben-Aryeh, Y., Mann, A.  \&
    Yaakov, I. Rabi oscillations in a two-level atomic system with a
    pseudo-hermitian hamiltonian. \emph{J. Phys. Math. Theor. } \textbf{ 37},  12059 (2004).}

\bibitem{Doppler2016}
{Doppler, J. et~al. Dynamically encircling an exceptional point for
    asymmetric mode switching.\emph{ Nature} \textbf{ 537} 76 (2016).}

\bibitem{JL16}
{Li, J., Liu, J., Xu, W., deMelo, L.  \&  Luo, L. Parametric cooling of a degenerate {Fermi} gas in an
    optical trap. \emph{Phys. Rev. A} \textbf{ 93},
    041401(R) (2016).}

\bibitem{ohara02a}
{O'Hara, K.~M. et~al. Measurement of the zero crossing in a \mbox{F}eshbach
    resonance of fermionic $^6$\mbox{L}i. \emph{Phys. Rev. A} \textbf{ 66},
    041401(R) (2002).}

\bibitem{KetterleZeno}
{Streed, E.~W. et~al. Continuous and pulsed quantum zeno effect.
\emph{Phys. Rev. Lett.} \textbf{97}, 260402 (2006).}

\bibitem{RaizenZeno}
{Fischer, M.~C., Gutierrez-Medina, B. \&
    Raizen, M.~G. Observation of the quantum zeno and anti-zeno effects
    in an unstable system. \emph{Phys. Rev. Lett.}
    \textbf{ 87},  040402 (2001).}

\bibitem{Zhu2014}
{Zhu, B. et~al. Suppressing the loss of ultracold molecules via the
    continuous quantum zeno effect. \emph{Phys. Rev. Lett.}
    \textbf{ 112},  070404 (2014).}

\bibitem{PatilZeno}
{Patil, Y.~S., Chakram, S.  \&
    Vengalattore, M. Measurement-induced localization of an ultracold
    lattice gas. \emph{Phys. Rev. Lett.}
    \textbf{ 115},  140402 (2015).}

\bibitem{Mailybaev2005}
{Mailybaev, A.~A., Kirillov, O.~N.  \&
    Seyranian, A.~P. Geometric phase around exceptional points. \emph{Phys. Rev. A}  \textbf{ 73},
    014104 (2005).}

\bibitem{Uzdin2011}
{Uzdin, R. , Mailybaev, A. \&
    Moiseyev, N. On the observability and asymmetry of adiabatic state
    flips generated by exceptional points.\emph{ J. Phys. Math. Theor.}
    \textbf{ 44},  435302
    (2011).}

\bibitem{Wu2014}
{Wu, J.-H., Artoni, M.  \&
    Rocca, G. C.~L. Non-hermitian degeneracies and unidirectional
    reflectionless atomic lattices. \emph{Phys. Rev. Lett.}
    \textbf{ 113},  123004 (2014).}

\bibitem{Gao2015}
{Gao, T. et~al. Observation of non-hermitian degeneracies in a
    chaotic exciton-polariton billiard. \emph{Nature}  \textbf{ 526},
    554 (2015).}

\bibitem{Milburn2015}
{Milburn, T.~J. et~al. General description of quasiadiabatic dynamical
    phenomena near exceptional points. \emph{Phys. Rev. A}  \textbf{ 92},
    052124 (2015).}

\bibitem{Hassan2017}
{Hassan, A.~U., Zhen, B., Soljacic, M., Khajavikhan, M. \&
    Christodoulides, D.~N. Dynamically encircling exceptional points: exact
    evolution and polarization state conversion. \emph{Phys. Rev. Lett.}
    \textbf{ 118},  093002 (2017).}

\bibitem{Yoon2018}
{Yoon, J.~W. et~al. Time-asymmetric loop around an exceptional point over
    the full optical communications band. \emph{Nature}  \textbf{ 562},
    86 (2018).}

\bibitem{Luo07Entropy}
{Luo, L., Clancy, B., Joseph, J., Kinast, J. \&  Thomas, J.~E.
    Measurement of the entropy and critical temperature of a strongly interacting \mbox{F}ermi gas.
    \emph{Phys. Rev. Lett.} \textbf{ 98},  080402 (2007).}


\end{thebibliography}

\section{Acknowledgements}
Le Luo is a member of the Indiana University Center for Spacetime
Symmetries (IUCSS). LL received supports from National Natural
Science Foundation of China under Grant No. 11774436, Sun Yat-sen
University Discipline Construction Fund, Sun Yat-sen University
Three Major Construction Fund, Indiana University Collaborative
Research Grant. YJ received NSF grant no. DMR-1054020. JLi received
supports from National Natural Science Foundation of China under Grant No. 11804406.

\section{Author contribution}
LL and YJ conceived the idea and supervised the study. JLi and LDm
set up experiments and performed measurements. LL designed and
supervised the experiments. AH, JLi, and YJ carried out theoretical
modeling. JLi and JLiu analyzed the data. JLi, LL, and YJ
contributed to writing the manuscript. Correspondence and requests
for materials should be addressed to YJ (yojoglek@iupui.edu) and LL
(luole5@mail.sysu.edu.cn).

\section{Competing interests:}
The authors declare no competing interests.

\section{Figures and Legends}

\begin{figure}[htbp]
\includegraphics[width=0.6\columnwidth]{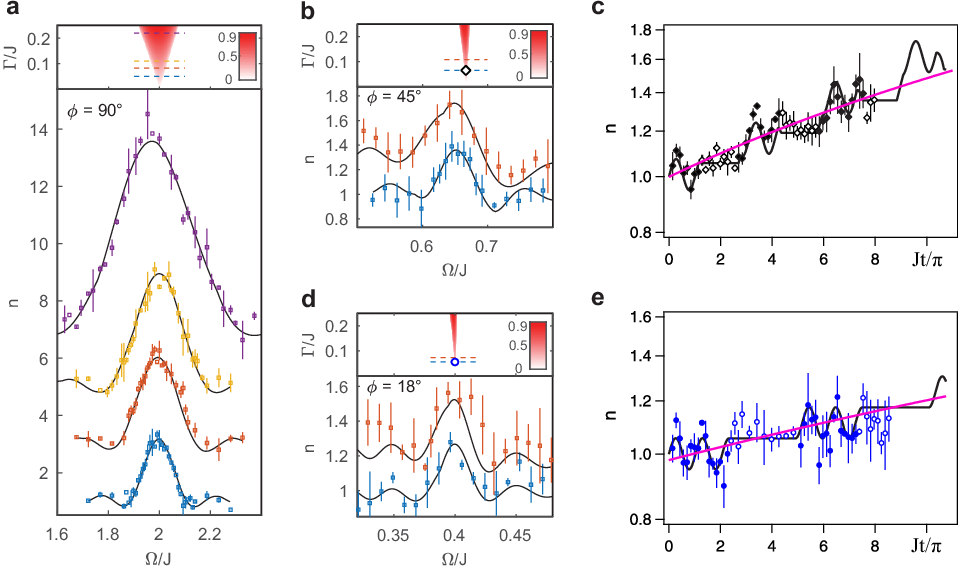}
\caption{Detecting the resonances with time-periodic dissipation.
\textbf{a} $n(t_{\text{f}},\Omega)$ around the one-photon resonance
$\Omega_{\text{n}}=2J$. Purple for $\Gamma/J$=0.22,
$t_{\text{f}}$=1.98 ms; Yellow for $\Gamma/J$=0.11,
$t_{\text{f}}$=3.31 ms; Red for $\Gamma/J$=0.082, $t_f$=3.97 ms; and
Blue for $\Gamma/J$=0.05, $t_{\text{f}}$=5.29 ms. \textbf{b}
$n(t_{\text{f}},\Omega)$ around the three-photon resonance
$\Omega_{\text{n}}=2J/3$. Red for $\Gamma/J=0.11$,
$t_{\text{f}}$=2.80 ms; Blue for $\Gamma/J=0.065$,
$t_{\text{f}}$=3.79 ms. \textbf{c} $n(t)$ at
$\Omega_{\text{n}}=2J/3$ with $\Gamma/J=0.065$ located in \textbf{b}
by the diamond shape. \textbf{d} $n(t_{\text{f}},\Omega)$ around the
five-photon resonance $\Omega_{\text{n}}=2J/5$. Red for
$\Gamma/J=0.07$, $t_{\text{f}}$=4.64 ms; Blue for $\Gamma/J=0.055$,
$t_{\text{f}}$=4.74 ms. \textbf{e} $n(t)$ at
$\Omega_{\text{n}}=2J/5$ with $\Gamma/J=0.055$ located in \textbf{d}
by the circle shape. The initial phase $\phi$ of the square-wave
modulation is chosen to anti-synchronize to the RF field with
$\phi=\pi/2,\pi/4,\pi/10$ for one-, three-, and five-photon
resonance respectively. In \textbf{a}, \textbf{b}, and \textbf{d},
all the data and simulation curves have a base line of $n=1$, but
are vertically shifted for the presentation purpose. The side peaks
are due to the finite probe time. In \textbf{c} and \textbf{e}, the
data with solid shapes corresponds with the dissipation on, the data
with empty shapes are with the dissipation off, and the pink curves
are the exponential fitting. For all figures, the solid curves are
numerical simulations without free parameter. The error bars
represents the standard deviation of the measurements.\label{fig2}}
\end{figure}

\begin{figure}[htbp]
\includegraphics[width=0.8\columnwidth]{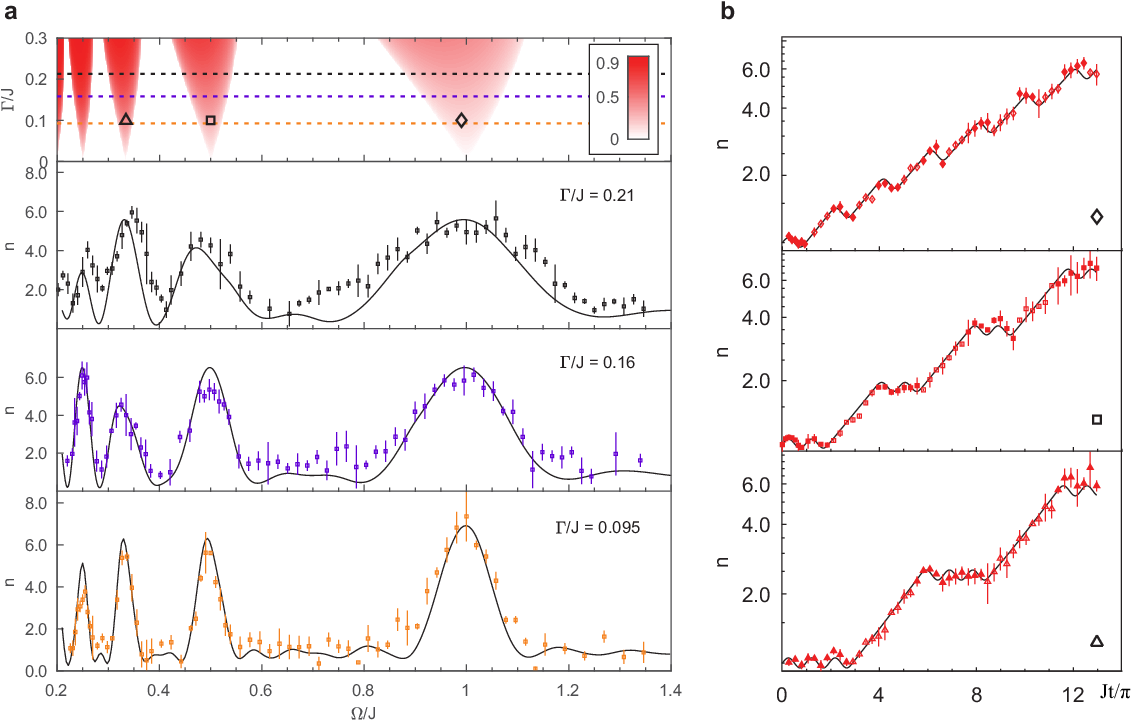}
\caption{Detecting multiphoton resonances with time-periodic
coupling. \textbf{a} The top frame is the phase diagram.
$n(t_{\text{f}},\Omega)$ are shown below. From the second top frame to the
bottom one, $\Gamma/J=0.21,\,0.16,\,0.095$, $t_{\text{f}}=2.70,\,3.70,\,6.48$
ms respectively. The side peaks are mainly due to the finite probe
time. \textbf{b} $n(t)$ with $\Gamma/J=0.095$ at the resonant peak
$\Omega_{\text{n}}=J,\,J/2,\,J/3$ from the top to bottom frame. The data with
solid shapes corresponds with the dissipation on, the data with
empty shapes are with the dissipation off. For all figures, solid
curves are the numerical simulation without free parameters. The
initial phases of the square-wave modulation are set to zero. The
error bars represent the standard deviation of the measurements.
\label{fig3}}
\end{figure}

\begin{figure}[h!]
\centering
\includegraphics[width=0.8\columnwidth]{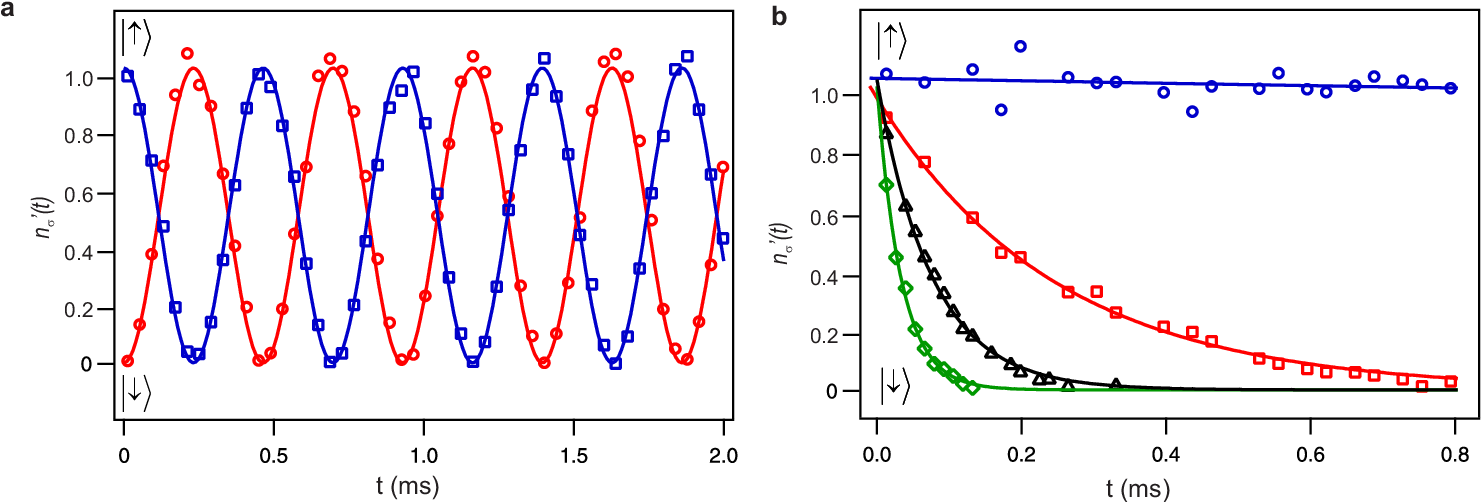}
\caption{Characterization of a dissipative Rabi system. Symbols are
experimental data. Solid lines are theoretical fits. \textbf{a} In
the presence of the RF field but without the dissipative optical
field, atom numbers $n'_{\sigma}(t)$ show Rabi oscillations with a
Rabi frequency of $2J=(2\pi)\times2.15$ kHz. \textbf{b} In the
presence of an optical field resonant with $\down\rangle$ but
without the RF field, $n'_\downarrow(t)$ shows an exponential decay:
$\Gamma=0.30J$ (red squares), $\Gamma=0.98 J$ (black triangles),
$\Gamma=2.35J$ (green diamonds). $n'_\uparrow(t)$ remains almost
constant during the experimental time. Blue circles show
$n'_\uparrow(t)$ when $\Gamma=0.30J$. \label{fig:sm12}}
\end{figure}

\end{document}